# Perfect spin-fillter and spin-valve in carbon atomic chains

M. G. Zeng,[1,2] L. Shen,[1] Y. Q. Cai,[1] Z. D. Sha,[1] and Y. P. Feng[1,*]

[1]Department of Physics, National University of Singapore, 2 Science Drive 3, Singapore 117542, Singapore
[2]Nanocore, 5A Engineering drive 4, National University of Singapore,117576, Singapore

We report *ab initio* calculations of spin-dependent transport in single atomic carbon chains bridging two zigzag graphene nanoribbon electrodes. Our calculations show that carbon atomic chains coupled to graphene electrodes are perfect spin-filters with almost 100 % spin polarization. Moreover, carbon atomic chains can also show a very large bias-dependent magnetoresistance up to $10^6$ % as perfect spin-valves. These two spin-related properties are independent on the length of carbon chains. Our report, the spin-filter and spin-valve are conserved in a single device simultaneously, opens a new way to the application of all-carbon composite spintronics.

Due to the ballistic quantum transport and remarkable long spin-coherence times and distances in carbon-based nanostructures, all-carbon nanodevices have attracted considerable attention for their possible application in electronics or spintronics.[1,2] Recently, all-carbon graphene nanoribbon (GNR) based field-effect transistors (FETs) are experimentally fabricated by Ponomarenko *et al.*, which make all-carbon electronics devices becoming realization.[3] In order to get semiconducting GNRFETs, sub-10-nm GNRs is necessary but difficult to get due to the limitation of the current lithography technique.[4] Very recently, linear carbon atomic chains have been carved out from graphene with a high energy electron beam in two groups.[5,6] Such ground-breaking experiments pave a wave for novel all-carbon FETs with many merits compared with GNR- or carbon nanotube- (CNT) based one. The reason is that carbon atomic chains are identical and can be considered as extremely narrow GNRs or thin CNTs. Therefore, they eliminate the need for sorting through a pile of different chiral GNRs and CNTs. Motivated by experiments, Shen *et al:*and Chen *et al:* theoretically studied the electron transport properties of carbon chain-graphene junctions and discussed their potential applications in electronics.[7,8]

Spintronics is an emerging technology that exploits the intrinsic spin degree of freedom of the electron. Several carbon-based materials are proposed for spintronics applications, such as graphene and carbon nanotubes. Graphene can be used as a spin valleytronics device by adjusting of its band valley[9] and Zigzag edged GNRs are predicted to be half-metallic under electrical field, which can be used as a spintronics device.[10] Tombros *et al:* experimentally studied spin-diffusion in graphene device and observed long spin flip time/length.[11] Kim *et al:* theoretically predicted a very large values of magnetoresistance in a GNR-based all-carbon FET as a spin valve.[12-14] Karpan *et al:* predicted graphene as a perfect spin filter when bridging ferromagnetic leads.[15] The narrowest GNRs, carbon atomic chains, also have been theoretically studied as perfect spin filters between nonmagnetic Au electrodes or spin-valves bridging Al electrodes under magnetic fields.[16,17] Moreover, modified carbon chains have been predicted as spin-filters or spin-valves. For example, Yang *et al:* proposed half-metallic properties of carbon nanowire inside a single-walled CNTs[18] and Senapati *et al:* predicted large magnetoresistance in Co-terminal carbon chains bridging Au electrodes.[19] Besides the metallic leads, GNRs or CNTs also can be used as leads in spintronics devices for spin injection.[20-24] Ke *et al:* and Koleini *et al:* proposed organometallic molecules as spin-filters when bridging CNT leads.[23,24] However, there is little report on spin-filter and spin-valve effect in a single spintronics device especially in carbonbased nanodevices.

In this letter, we explore spin-dependent electron transport in carbon chain bridging two zigzag GNR electrodes (see Fig. 1). When spin electrons inject in carbon chains from metallic GNR leads, the majority-spin (α) transport channel is fully open while the minority-spin (β) channel is blocked. The carriers are thus 100 % polarized which is the ideal case for spin filters. Moreover, we study magnetoresistance (MR) by changing the spin orientations of two leads. Very large values of MR up to $10^6$ % are observed in the same model. We change the length of carbon chains from 3 to 16 atoms and find that these two novel spin-related properties are independent on the length of carbon chains.
Our first-principles calculations are based on spin density functional theory combined with nonequilibruim Green's function as implemented in the ATK package[25,26]. The mesh cutoff is 150 Ry and Monkhorst-Pack sampling of $1\times1\times100$. A duoble-ξ polarized (DZP) basis set is used in order to preserve a correct description of π conjugated bonds. The models are optimized as in Ref. [7]. Zigzag GNRs are used as leads due to their perfect conductivity. The vacuum layers between two sheets along

$z$ and $x$ directions are 19 Å in order to eliminate the effect of neighboring cells (see Fig. 1).

We first calculate the spin-dependent electron transmission in $C_7$ and $C_8$ as plotted in Fig. 2(a)-(d). It can be seen that there is no spin-polarization between α state and β state both in $C_7$ and $C_8$ if the spin orientation of two leads is antiparallel alignment. Moreover, both α and β channels are blocked at the Fermi level. In the other hand, spin-polarization occurs with a large spin splitting energy of 450 and 345 $meV$ in $C_7$ and $C_8$, respectively with parallel aligned spin current injection. In the parallel configuration, the spin polarization of the electron current both approaches 100 % based on the equation:

$$TSP = (T_{P\text{-}\alpha} - T_{P\text{-}\beta})/(T_{P\text{-}\alpha} + T_{P\text{-}\beta}) \times 100\%$$

where $TSP$ is the transmission spin polarization. The different types of conduction mechanisms in the two spin channels are reflected in the space-resolved local density of states (LDOS) at the Fermi level. The LDOS of the $C_8$ case are plotted in the lower panel of Fig.2. As can be seen, the carbon atomic chains are perfect spin-filters.

Next, we calculate the spin-resolved current-voltage ($I$-$V$) characteristics from the bias-dependent transmission curves using the Landauer-Buttiker formula:[12,13]

$$I(V_b) = e/h \int T(E, V_b) [f_L(E, V_b) \times f_R(E, V_b)] dE$$

where $e$, $h$ and $f_{L(R)}(E, V_b)$ are the electron charge, Planck's constant, and the Fermi distribution functions at left (right) electrode, respectively. $T(E; V_b)$ is the transmission coefficient at energy $E$ and bias voltage $V_b$. Figure 3 shows the $I$ - $V$ curve of the $C_7$ case. As can be seen, the majority-spin state of the parallel configuration is metallic while the minority-spin state is insulating. For the antiparallel configuration, it shows semiconducting with the threshold voltage of 100 $mV$ because of the low value of the transmission below the threshold voltage (see Fig. 2(a)). Magnetoresistances can be obtained from the $I$ - $V$ curves using the following equation:

$$MR = [R_{AP} - R_P] / R_P \times 100\% = [(T_{AP} + T_{AP}) - (T_P + T_P)] / (T_P + T_P) \times 100\%$$

where $R_{AP}$ and $R_P$ are the resistances in the antiparallel and parallel configuration of two leads. The inset of Fig. 3 shows the bias-dependent MR of carbon chains with seven carbon atoms. The MR shows an decay function of bias voltage and the maximum value at zero bias is large than $10^5$ %. It still holds a large value ($10^3$ %) under a small bias, which is an order larger than that of conventional spin-valve devices ($10^2$ %).[28,29] This is because that carbon chains have more selective transmission with the additional orbital matching to graphene nanoribbon leads compared with conventional devices. The very large values of MR in carbon chain FETs make these devices having the potential in achieving ideal spin-valve devices.

Finally, we investigate the spin-dependent transmission on different length of carbon atomic chains from three to sixteen carbon atoms in Table 1. As can be seen, the large TSP and MR of carbon atomic chains are independent on the length of carbon chains. But the longer carbon chains have more stable values than the shorter one. For example, the TSP keeps constancy of 100% and the MRs have a small fluctuation between $1.4 \times 10^5$ ~ $2.4 \times 10^6$ %, when the chains are longer than the one with seven carbon atoms. Almost 100 % TSP indicates the perfect spin-filter effect and very large values of MR indicates a highly efficient spin-valve effect. Note that the values of $T_P$ have an odd-even oscillatory property. It is due to the bond-length alternation of carbon chains with odd and even-numbered carbon atoms after fully structural relaxation.[6,7]

In conclusion, motivated by recent experiments we systematically study the spin-dependent transport in carbon atomic chain-based all-carbon FETs. These devices could generate perfect spin-polarized currents and serve as highly effective spin-filters. When we alternate the spin orientation of two metallic leads, the spin-polarized current will be on-off consequently as good spin-valves.
These two interesting spin-related properties in carbon chain-graphene FETs hold the promise of all-carbon composite devices for spintronics.

During the work of this manuscript, we became aware of a recent job of Furst *et al*,[27] who proposed a complete spin-polarization of the transmission in large energy ranges of carbon chains connecting two semi-infinite graphene sheets at the zigzag edges. Under a small bias, their devices could be perfect spin-filters. Our calculation, using zigzag GNRs as leads, shows a similar spinpolarization in carbon chain-graphene junctions, which could be perfect spin-filters without additional bias due to the large spin-splitting at the Fermi level.

The authors thank Dr. Xu Bo and Dr. Wu Rongqing for their helpful discussion.

Electronic address: phyfyp@nus.edu.sg

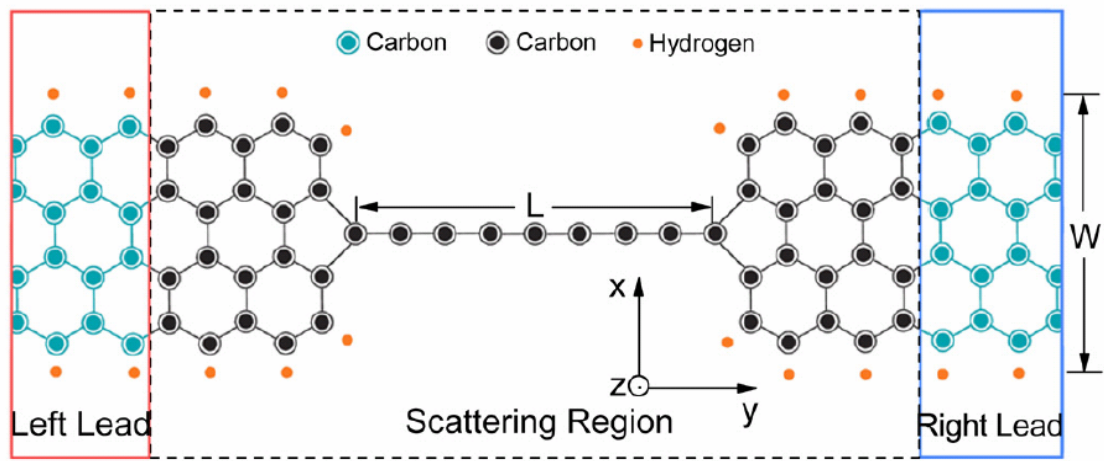

FIG. 1: (color online) A schematic device model of carbon atomic chains bridging two zigzag graphene nanoribbon leads. The red frame indicates the left leads while the blue one indicates the right leads. The scattering region includes the carbon chain and the surface layers. The direction of electron transport is along the carbon atomic chain labeled as *y*.

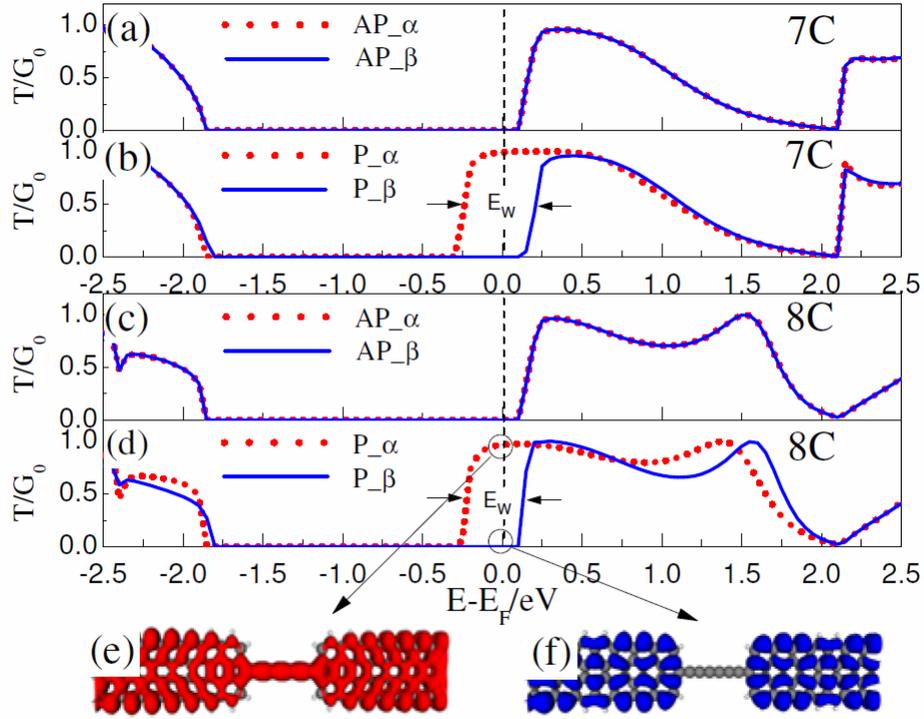

FIG. 2: (color online) The spin-dependent electron transmission at zero bias. (a)-(b) Spin transmission of $C_7$ with the antiparallel/parallel spin orientation of two leads. (c)-(d) Spin transmission of $C_8$ with the antiparallel/parallel spin orientation of two leads. (e)-(f) show surfaces of the constant spinresolved local DOS evaluated at the Fermi level.

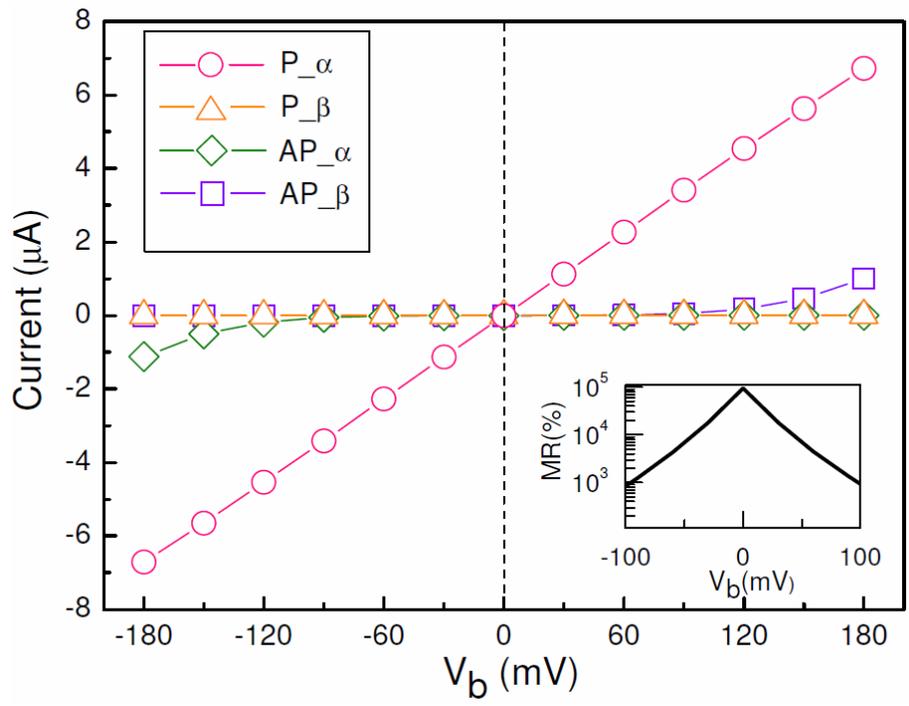

FIG. 3: (color online) The spin-resolved I-V curve of $C_7$ with the parallel/antiparallel spin orientation of two leads. The inset is bias-voltage dependent magnetoresistance.

TABLE I: Based on the spin dependent transmission $T$ ($e^2/h$) with antiparallel ($AP$) and parallel ($P$) configuration of different length ($L$) of carbon chains, the transmission spin polarization ($TSP$) and magnetoresistance ($MR$) are calculated.

| L | Antiparallel configuration | | Parallel configuration | | $TSP$ (%) | $MR$ (%) |
|---|---|---|---|---|---|---|
| | $T_{AP-\alpha}$ | $T_{AP-\beta}$ | $T_{P-\alpha}$ | $T_{P-\beta}$ | | |
| $C_3$ | $2.86\times10^{-5}$ | $5.46\times10^{-5}$ | 0.88 | $4.56\times10^{-5}$ | 100 | $1.05\times10^6$ |
| $C_4$ | $5.90\times10^{-3}$ | $5.87\times10^{-3}$ | 0.85 | $5.12\times10^{-4}$ | 99.88 | $7.15\times10^3$ |
| $C_5$ | $5.43\times10^{-3}$ | $5.61\times10^{-3}$ | 0.89 | $2.56\times10^{-5}$ | 100 | $7.92\times10^3$ |
| $C_6$ | $7.17\times10^{-3}$ | $7.25\times10^{-3}$ | 0.91 | $1.79\times10^{-3}$ | 99.61 | $6.25\times10^3$ |
| $C_7$ | $4.02\times10^{-4}$ | $2.71\times10^{-4}$ | 0.97 | $1.45\times10^{-8}$ | 100 | $1.44\times10^5$ |
| $C_8$ | $2.13\times10^{-5}$ | $1.69\times10^{-5}$ | 0.82 | $1.66\times10^{-9}$ | 100 | $2.14\times10^6$ |
| $C_9$ | $3.22\times10^{-5}$ | $6.40\times10^{-5}$ | 0.98 | $9.40\times10^{-10}$ | 100 | $1.02\times10^6$ |
| $C_{10}$ | $1.38\times10^{-5}$ | $1.88\times10^{-5}$ | 0.81 | $6.10\times10^{-10}$ | 100 | $2.47\times10^6$ |
| $C_{11}$ | $9.17\times10^{-5}$ | $1.49\times10^{-5}$ | 0.99 | $1.72\times10^{-9}$ | 100 | $4.12\times10^5$ |
| $C_{12}$ | $1.41\times10^{-5}$ | $2.07\times10^{-5}$ | 0.79 | $1.28\times10^{-9}$ | 100 | $2.26\times10^6$ |
| $C_{13}$ | $8.03\times10^{-5}$ | $6.98\times10^{-5}$ | 0.91 | $1.46\times10^{-9}$ | 100 | $6.01\times10^5$ |
| $C_{14}$ | $2.25\times10^{-5}$ | $7.80\times10^{-5}$ | 0.78 | $1.28\times10^{-9}$ | 100 | $2.56\times10^6$ |
| $C_{15}$ | $2.16\times10^{-5}$ | $6.60\times10^{-5}$ | 1.00 | $7.28\times10^{-10}$ | 100 | $3.54\times10^6$ |
| $C_{16}$ | $5.19\times10^{-5}$ | $2.51\times10^{-5}$ | 0.75 | $3.58\times10^{-9}$ | 100 | $9.77\times10^5$ |